\documentclass[aps,prl,twocolumn,superscriptaddress,showpacs,floatfix]{revtex4-1}

\usepackage{hyperref}
\usepackage{color}
\usepackage{graphicx}	
\usepackage[utf8]{inputenc}
\usepackage{amsmath}
\usepackage{amssymb}

\begin{document}
\title{Holographic Heavy-Ion Collisions: Analytic Solutions with Longitudinal Flow, Elliptic Flow and Vorticity}

\author{Hans Bantilan}
\affiliation{School of Mathematical Sciences, Queen Mary University of London, E1 4NS, UK}
\affiliation{Centre for Research in String Theory, School of Physics and Astronomy, Queen Mary University of London, E1 4NS, UK}
\affiliation{DAMTP, University of Cambridge, CB3 0WA, UK}
\author{Takaaki Ishii}
\affiliation{Institute for Theoretical Physics, Utrecht University, 3584 CC Utrecht, The Netherlands}
\author{Paul Romatschke}
\affiliation{Department of Physics, University of Colorado, Boulder, Colorado 80309, USA}
\affiliation{Center for Theory of Quantum Matter, University of Colorado, Boulder, Colorado 80309, USA}

\begin{abstract}
  We consider phenomenological consequences arising from simple analytic solutions for holographic heavy-ion collisions. For these solutions, early-time longitudinal flow is initially negative (inward), sizable direct, elliptic, and quadrangular flow is generated, and the average vorticity of the system is tunable by a single parameter. Despite large vorticity and angular momentum, we show that the system does not complete a single rotation.
  \end{abstract}

\maketitle
\textbf{Introduction}
%
%
%
The theoretical description of heavy-ion collisions, such as Pb+Pb collisions at $\sqrt{s}=5.02$ TeV at the Large Hadron Collider (LHC)\cite{Schukraft:2017nbn,Busza:2018rrf}, is a difficult problem, given that it combines real-time dynamics of QCD at interaction strengths of order unity. At present, the infamous sign problem inhibits the use of lattice QCD techniques to calculate real-time dynamics, whereas the strong coupling nature of the problem renders the application of perturbative techniques questionable at best. One tool that is available to calculate real-time dynamics in quantum field theories in the limit of large number of colors and strong coupling is the conjectured duality between gauge theories and gravity \cite{Maldacena:1997re}. Within gauge/gravity duality, the collision of two heavy-ions may be recast as the problem of black hole collisions in classical gravity in five-dimensional asymptotic anti-de-Sitter (AdS) space-time \cite{Nastase:2005rp,Amsel:2007cw,Grumiller:2008va,Gubser:2008pc,Albacete:2008vs}, for which numerical solutions exist \cite{Bantilan:2014sra} (see also \cite{Chesler:2010bi,Heller:2011ju,vanderSchee:2012qj,Casalderrey-Solana:2013aba,vanderSchee:2013pia,Fernandez:2014fua,Chesler:2015wra,Ecker:2016thn}). The gravitational dual description is not an exact map of the real-world collision problem, given that the gravitational dual to QCD is not known, and no proof for gauge/gravity duality itself exists to date. Therefore, at best gauge/gravity techniques currently are ``only'' able to exactly solve the problem of the collision of matter in strongly coupled ${\cal N}=4$ supersymmetric Yang-Mills (SYM) theory (as well as similar gauge theories not realized in nature) in the limit of large number of colors and strong coupling. While this limitation precludes the calculation of \textit{quantitatively} accurate results for real-world heavy-ion collision, the ability to obtain rigorous solutions of the real-time dynamics of strongly coupled gauge theories from first principles opens up the possibility for novel \textit{qualitative} or even semi-quantitative insights which are currently unattainable with traditional weak-coupling techniques. One of the more famous success stories of this approach is the prediction for the shear viscosity over entropy ratio from gauge/gravity duality \cite{Policastro:2001yc}, which is within a factor of two of the value extracted from comparisons between hydrodynamic model calculations and heavy ion collision data \cite{Romatschke:2007mq,Gale:2012rq,Bernhard:2016tnd}. Other examples include the realization (originally based on exact results in gauge/gravity duality) that the onset of hydrodynamic behavior following a heavy-ion collision is unrelated to the thermalization of the system, arising instead from the decay of so-called non-hydrodynamic modes \cite{CasalderreySolana:2011us,Keegan:2015avk,Heller:2016rtz,Romatschke:2016hle,Heller:2018qvh}, as well as the notion that a non-perturbative formulation of hydrodynamics, sometimes referred to as hydrodynamic attractors \cite{Lublinsky:2007mm,Heller:2015dha,Romatschke:2017vte,Spalinski:2017mel,Strickland:2017kux,Romatschke:2017acs,Denicol:2017lxn,Behtash:2017wqg,Casalderrey-Solana:2017zyh}, allow quantitatively accurate descriptions of systems below the femtoscale (see Refs.~\cite{Florkowski:2017olj,Romatschke:2017ejr} for recent reviews). Taken together, these realizations provide a firm theoretical foundation for the otherwise ``unreasonable success'' of hydrodynamic descriptions of high energy proton-proton collisions \cite{Weller:2017tsr}.

\begin{figure*}[t]
  \hspace*{-2cm}
  \includegraphics[width=0.47\linewidth]{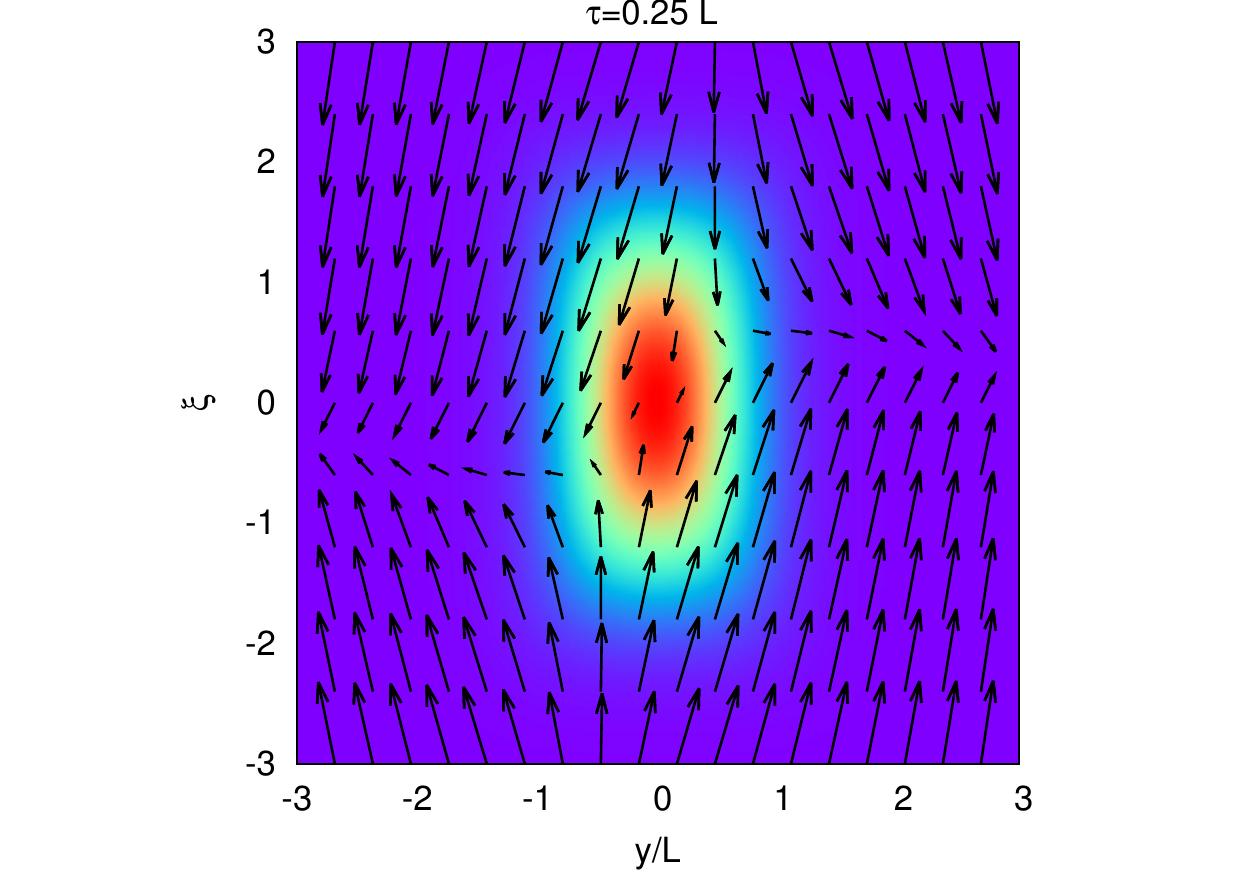}\hspace*{-2cm}
  \includegraphics[width=0.47\linewidth]{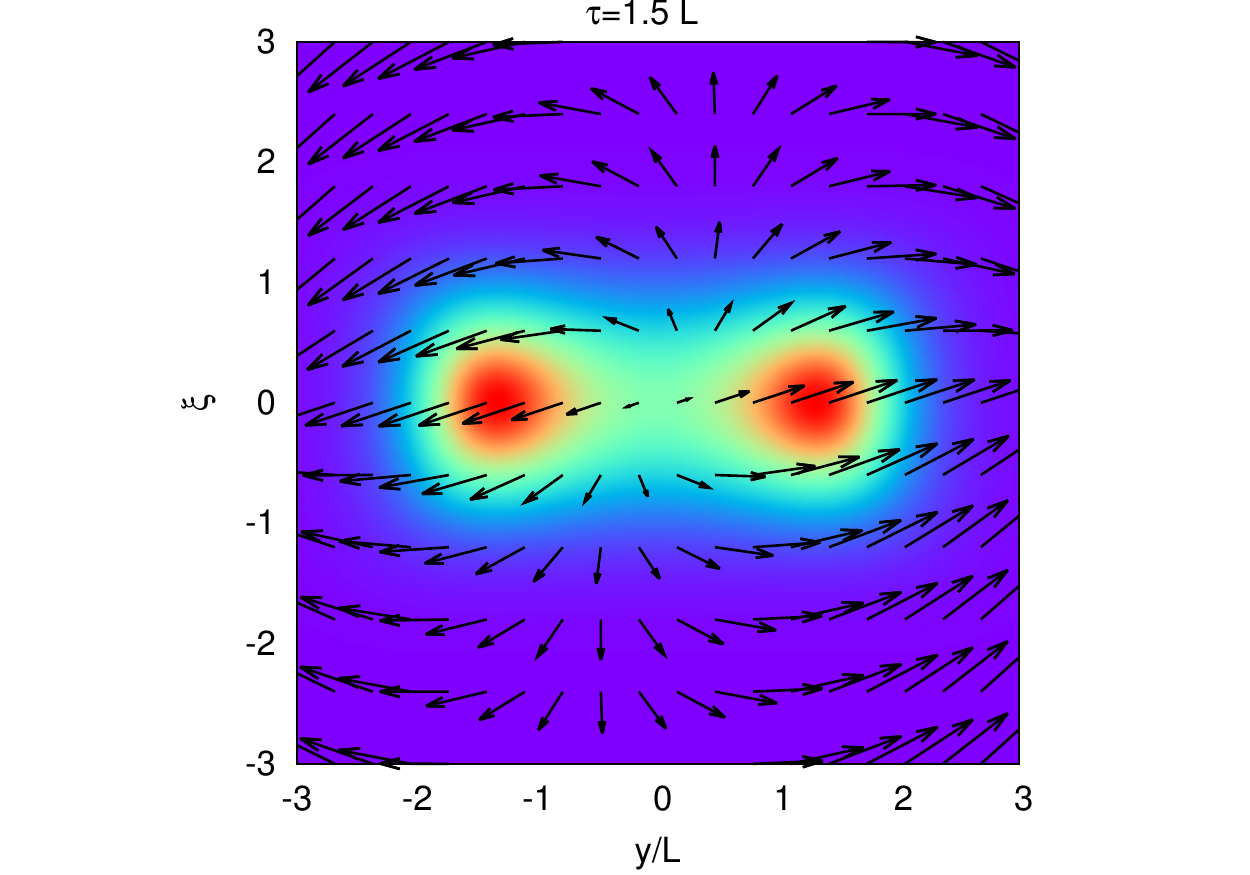}\hspace*{-2cm}
  \includegraphics[width=0.47\linewidth]{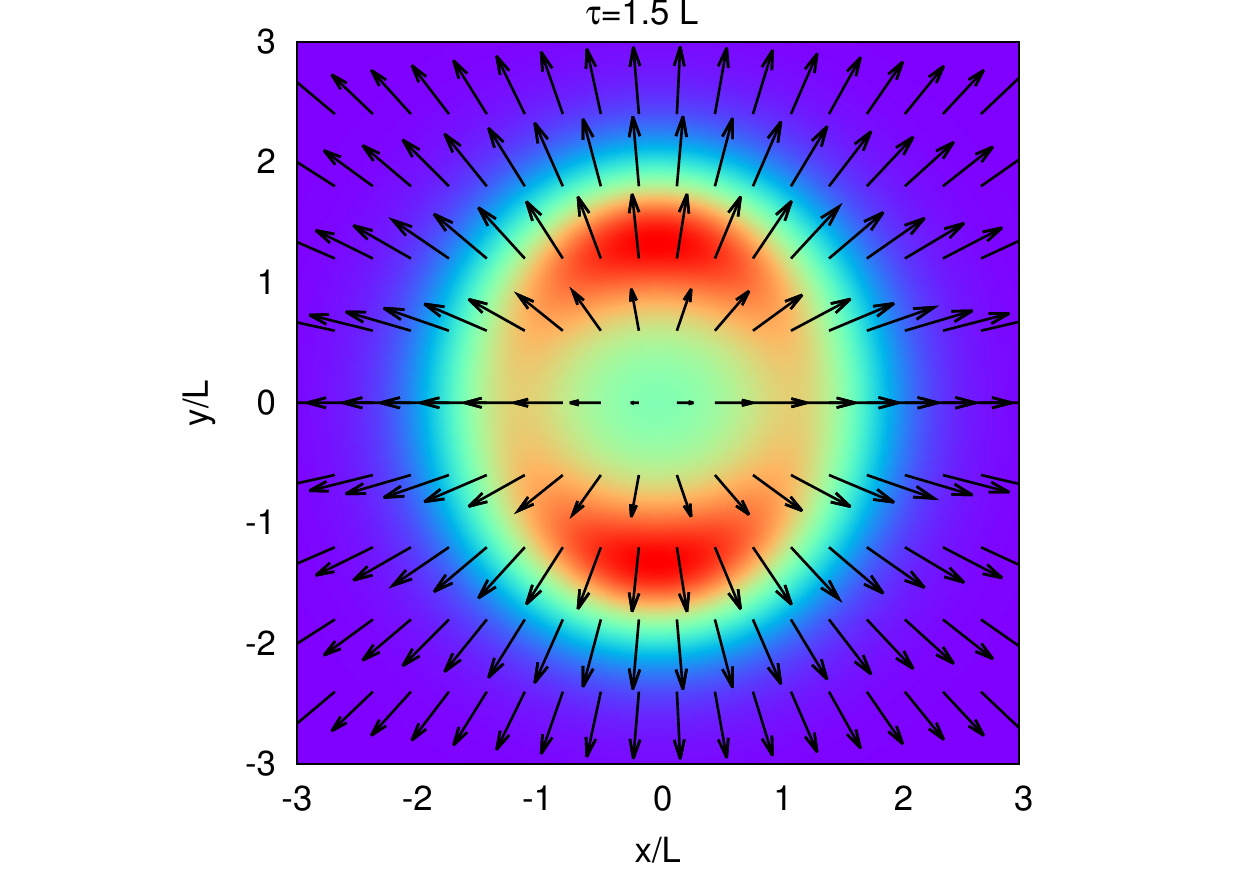}
  \caption{\label{fig:flow1} Flow velocity (vectors) for $\omega_1=0.5$, $\omega_2=0.05$ in the $y-\xi$ plane (left and middle panels with $x=0$) and the transverse plane (right panel with $\xi=0$). Also shown are iso-contours for the energy density. One can clearly see the negative longitudinal flow at early times and large $|{\bf x}_\perp|$ as well as the non-vanishing vorticity in the $y-\xi$ plane. The non-vanishing elliptic flow in the transverse plane is somewhat visible in the right panel.}
\end{figure*}

\textbf{Insights from Numerical Simulations}
The prospect of obtaining qualitative guidance for heavy-ion collisions provides the motivation for the study of the real-time evolution of ${\cal N}=4$ SYM matter arising from black hole collisions in AdS space-times. In particular, a gravity dual exhibiting so-called elliptic flow, an ubiquitous feature in real-world heavy-ion collisions, has so far remained elusive, because either only head-on collisions were simulated \cite{Bantilan:2014sra} or initial conditions with negligible spatial eccentricity were chosen \cite{Chesler:2015wra}. A key difference in the available numerical setups is the choice of coordinates: while Ref.~\cite{Chesler:2015wra} used the Poincar\'e patch of AdS with Minkowski space $\mathbb{R}^{3,1}$ as a boundary, simulations in Ref.~\cite{Bantilan:2014sra} were performed in global AdS having $S^3\times \mathbb{R}^1$ as boundary, together with simple coordinate and conformal transformations to obtain boundary data for Minkowski space. A key point from the simulations of Ref.~\cite{Bantilan:2014sra} is that the single (deformed) AdS-Schwarzschild black hole which results from a head-on black hole collision can be viewed in the Poincare patch as a black hole that gradually falls towards the Poincare horizon (see Ref.~\cite{Friess:2006kw} for a discussion), and thus corresponds to an expanding lump of energy density on a Minkowski piece of the boundary. Hence, while the actual collision process is complicated for either coordinate choice, results in Ref.~\cite{Bantilan:2014sra} demonstrated that the time-dependent problem after the collision corresponds to the ring-down of a single Schwarzschild black hole in the center of global AdS$_5$ for which the late-time limit is known analytically \cite{Friess:2006kw}.

The situation can be generalized to the case of off-central collisions of two black holes in global AdS$_5$. Even without numerically simulating the collision dynamics, the result must lead to the ring-down of a single Myers-Perry black hole in AdS$_5$ unless the black holes miss each other and no common horizon forms. Similarly, one may consider the case of colliding charged black holes, leading to the ring-down of a Reissner-Nordstr\"om black hole, and in the most general case to a Kerr-Newman black hole in global AdS$_5$.

\textbf{Analytic Hydrodynamic Solutions for Off-Center Heavy-Ion Collisions}
While detailed numerical studies are needed to accurately capture the ring-down dynamics of the deformed black hole, the dynamics becomes simple after the non-hydrodynamic quasinormal modes have decayed, because then the black hole becomes stationary in global AdS. Because stationarity precludes dissipative effects, these black hole solutions in global AdS correspond to solutions of ideal conformal hydrodynamics for the boundary gauge theory on $S^3\times \mathbb{R}^1$, such as those presented in Ref.~\cite{Bhattacharyya:2007vs},\footnote{Note that the holographic solutions actually correspond to solutions of hydrodynamics including arbitrary high order gradient corrections. Since the shear stress vanishes identically to all orders in gradients, none of these gradient corrections contributes, and the solution therefore is a solution of the ideal hydrodynamic equations of motion.}.

For the present work, we will consider the case of uncharged rotating black holes in global AdS (global Myers-Perry-AdS), corresponding to the case of an off-center black hole collision in global AdS$_5$ space-time. The generalization of the Kerr metric in dimensions higher than four, known as a Myers-Perry black hole, was first written down in \cite{Myers:1986un}, and the generalization that includes a negative cosmological constant was obtained in Ref.~\cite{Hawking:1998kw}, and these were generalized to all dimensions in Ref.~\cite{Gibbons:2004js}.  Using $\epsilon$ to denote the local energy density and $u^\mu=\gamma(1,v^x,v^y,v^\xi/\tau)$ to denote the fluid four-velocity, the ideal relativistic hydrodynamic solution on $S^3\times \mathbb{R}^1$ may be transformed to Minkowski space-time in Milne coordinates $\tau=\sqrt{t^2-z^2}$, $x$, $y$, $\xi=\mathrm{arctanh}(z/t)$ with metric tensor $g_{\mu\nu}={\rm diag}(-1,0,0,\tau^2)$ as
\begin{eqnarray}
  \label{eq:oursol}
   \epsilon &=&16 L^8 T_0^4 \left[(L^4+2 L^2 {\bf x}_\perp^2+(\tau^2-{\bf x}_\perp^2)^2)(1-\omega_2^2)\right.\nonumber\\
   &&\left.+2 L^2 (\tau^2-2 y^2)(\omega_1^2-\omega_2^2)+2 L^2 \tau^2(1-\omega_1^2)\cosh 2\xi\right]^{-2}\,,\nonumber\\
\gamma&=&\frac{\left[(L^2+\tau^2+{\bf x}_\perp^2)\cosh \xi+2 (\tau \omega_2 x-L \omega_1 y\sinh\xi)\right]}{\left(16 L^8 T_0^4/\epsilon\right)^{1/4}} \nonumber\\
v^x&=&\frac{2 \tau x \cosh\xi+\omega_2(L^2+\tau^2+x^2-y^2)}{(L^2+\tau^2+{\bf x}_\perp^2)\cosh \xi+2 (\tau \omega_2 x-L \omega_1 y \sinh\xi)}\,,\nonumber\\
  v^y&=&\frac{2 \tau y \cosh\xi+2 \omega_2 x y-2 L \tau \omega_1 \sinh\xi}{(L^2+\tau^2+{\bf x}_\perp^2)\cosh \xi+2 (\tau \omega_2 x-L \omega_1 y \sinh\xi)}\,,\nonumber\\
v^\xi&=&-\frac{(L^2-\tau^2+{\bf x}_\perp^2)\sinh\xi-2 L \omega_1 y \cosh\xi}{(L^2+\tau^2+{\bf x}_\perp^2)\cosh \xi+2 (\tau \omega_2 x-L \omega_1 y \sinh\xi)}\,,
  \end{eqnarray}
where ${\bf x}_\perp^2=x^2+y^2$, $T_0$ denotes the overall energy scale, $L$ is the AdS length scale that corresponds to a choice of units and $|\omega_{1,2}|<1$ are two angular rotation frequencies with $\omega_{1,2}=0$ corresponding to the case of no rotation and $\omega_{1,2}=\pm1$ corresponding to a Myers-Perry-AdS black hole rotating at the mass-shedding limit (see the Supplemental Material for details on how to obtain the solution (\ref{eq:oursol}) using Refs.~\cite{Hawking:1998kw,Hawking:1999dp,Gibbons:2004uw,Bhattacharyya:2007vs,Bantilan:2012vu}). Denoting the geometric covariant derivative as $\nabla_\mu$ and introducing $\nabla_\mu^\perp\equiv \nabla_\mu+u_\mu u^\nu \nabla_\nu$, one may verify that Eq.~(\ref{eq:oursol}) fulfills the ideal relativistic fluid dynamics equations of motion $u^\mu \nabla_\mu \epsilon=-(\epsilon+P) \nabla_\mu u^\mu$, $(\epsilon+P)u^\mu \nabla_\mu u^\alpha =-\nabla^\alpha_\perp P$ where  for a conformal fluid $P=\epsilon/3$.

\begin{figure*}[t]
  \hspace*{-2cm}
  \includegraphics[width=0.47\linewidth]{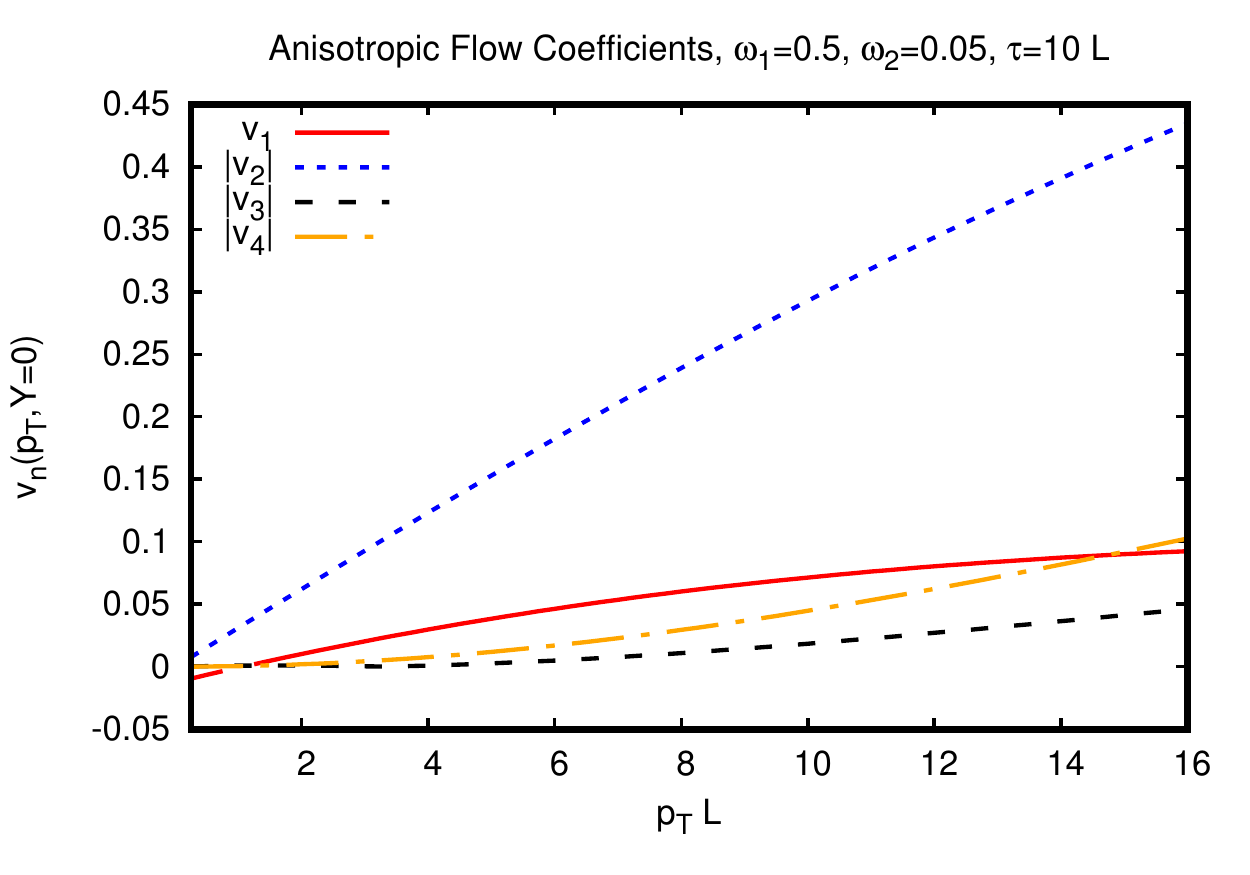}
  \includegraphics[width=0.47\linewidth]{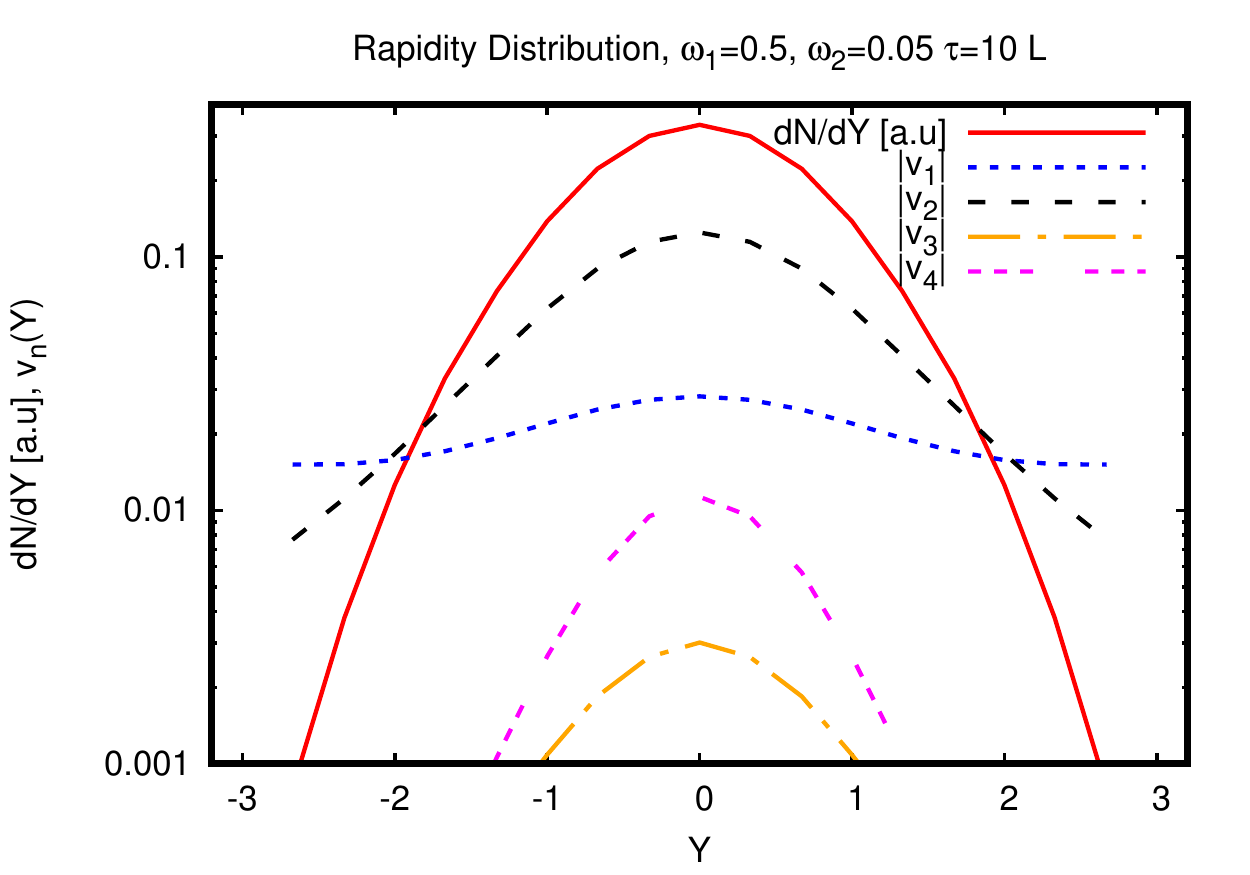}
  \caption{\label{fig:coeff} Left: Anisotropic Flow coefficients $v_{n}(p_T,Y=0)$ at mid-rapidity for massless particles as a function of transverse momentum $p_\perp$. Right: Multiplicity $\frac{dN}{dY}$ and $p_\perp$ integrated flow coefficients $v_n(Y)$ as a function of rapidity (see text for details). Both panels are for isochronous decoupling at $\tau_f=10 L$, but results are essentially insensitive to the choice of $\tau_f$. }
\end{figure*}

Analytic solutions to relativistic ideal hydrodynamics are often useful to test numerical solvers or to gain physical intuition. In particular, exact solutions for the Hubble flow in cosmology \cite{Weinberg:2008zzc}, and Bjorken \cite{Bjorken:1982qr} and Gubser flow \cite{Gubser:2010ze} in high energy nuclear collisions have proven to be particularly important. Eq.~(\ref{eq:oursol}) is a new analytic solution of relativistic ideal hydrodynamics that corresponds to a generalization of Refs.~\cite{Nagy:2009eq,Hatta:2014gga} to the case of rotation around two axes, and as such is similar to other exact analytic solutions that have been discussed in the literature \cite{Biro:2000nj,Csorgo:2003rt,Csorgo:2003ry,Csorgo:2006ax,Friess:2006kw,Bialas:2007iu,Beuf:2008vd,Peschanski:2009tg,Fouxon:2008ik,Liao:2009zg,Lin:2009kv}. However, Eq.~(\ref{eq:oursol}) has the attractive features that it arises naturally in the context of off-center black hole collisions in global AdS, that it has no apparent singularities for $\tau>0$ and $|\omega_{1,2}|<1$, and that it contains key features relevant to the phenomenology of heavy-ion collisions such as three dimensional evolution, longitudinal flow, elliptic flow, triangular flow and vorticity (cf. Fig.~\ref{fig:flow1}). In Eq.~(\ref{eq:oursol}), $\omega_1$ can be recognized to parametrize rotations in the $y-\xi$ plane, e.g. arising from non-vanishing angular momentum for off-center collisions, while $\omega_2$ parametrizes asymmetries in the x-y plane, which may e.g. arise from inhomogeneities of incoming nuclei (or the spin of colliding black holes in the dual gravitational description), suggesting that $|\omega_2|\ll 1$ would be appropriate for phenomenology. 

An immediate consequence from Eq.~(\ref{eq:oursol}) is that for forward space-time rapidities $\xi>0$, the longitudinal flow velocity is \textit{negative} for early times. ``Negative'' longitudinal flow implies that the system is collapsing toward mid-rapidity initially, rather than expanding, and to this extent Eq.~(\ref{eq:oursol}) may be interpreted as possessing some memory from the collision process itself, even if the flow is hydrodynamic. This behavior seems to be generic in holographic collisions, and has in particular also been found in Ref.~\cite{Grumiller:2008va}. There are hints that this phenomenon is present in heavy-ion experimental data, see Ref.~\cite{Stephanov:2014hfa}.

\textbf{Phenomenology}
Experimental detectors measure particles, not velocity fields. To convert the information from the analytic hydrodynamic solution (\ref{eq:oursol}) to particle information we employ the standard Cooper-Frye decoupling procedure \cite{Cooper:1974mv} that relates the fluid energy-momentum tensor to that of weakly interacting hadrons, leading to the particle spectrum for a hadron species $i$ given by
\begin{equation}
  \label{eq:fo}
  \frac{dN_i}{d^2 p_\perp dY}=- d_i \int \frac{p^\mu d\Sigma_\mu}{(2 \pi)^3} f_{\rm eq}(p^\mu u_\mu/T)\,,
\end{equation}
where $N_i$ is the number of hadrons of species $i$ with mass $m_i$, rapidity $Y$, transverse momentum $p_\perp=\sqrt{p_x^2+p_y^2}$ and spin/isospin degeneracy factor $d_i$ and $p^\mu$ is the particle's four-momentum. Here $d\Sigma_\mu$ is the normal vector on the decoupling hypersurface $\Sigma$ (which has to be defined by a criterion such as constant temperature $T=\epsilon^{1/4}$ or constant proper time $\tau$) and $f_{\rm eq}$ is the equilibrium particle distribution function (dissipative corrections to Eq.~(\ref{eq:fo}), which are reviewed e.g. in Ref.~\cite{Romatschke:2017ejr}, are absent for ideal hydrodynamics). Aiming for qualitative insight, we employ classical statistics $f_{\rm eq}(x)=e^{x}$ and isochronous decoupling for which $-p^\mu d\Sigma_\mu p^\mu=\tau \sqrt{m_i^2+p_\perp^2} \cosh(Y-\xi) d^2{\bf x}_\perp d\xi$. Decomposing the particle spectrum (\ref{eq:fo}) in terms of transverse Fourier components one obtains the anisotropic flow coefficients $v_n(p_\perp,Y)$ and flow angles $\psi_n(p_\perp,Y)$ as
$  \frac{dN_i}{d^2 p_\perp dY}=\frac{dN_i}{2\pi p_\perp dp_\perp dY}\left(1+2 \sum_{n=1}^\infty v_n(p_\perp,Y)
\cos\left[n (\phi-\psi_n(p_\perp,Y))\right]\right)$, where $p_{x}=p_\perp\cos\phi$ \cite{Ollitrault:1992bk,Voloshin:1994mz}.

%
%
\begin{figure*}[t]
  \hspace*{-2cm}
  \includegraphics[width=0.47\linewidth]{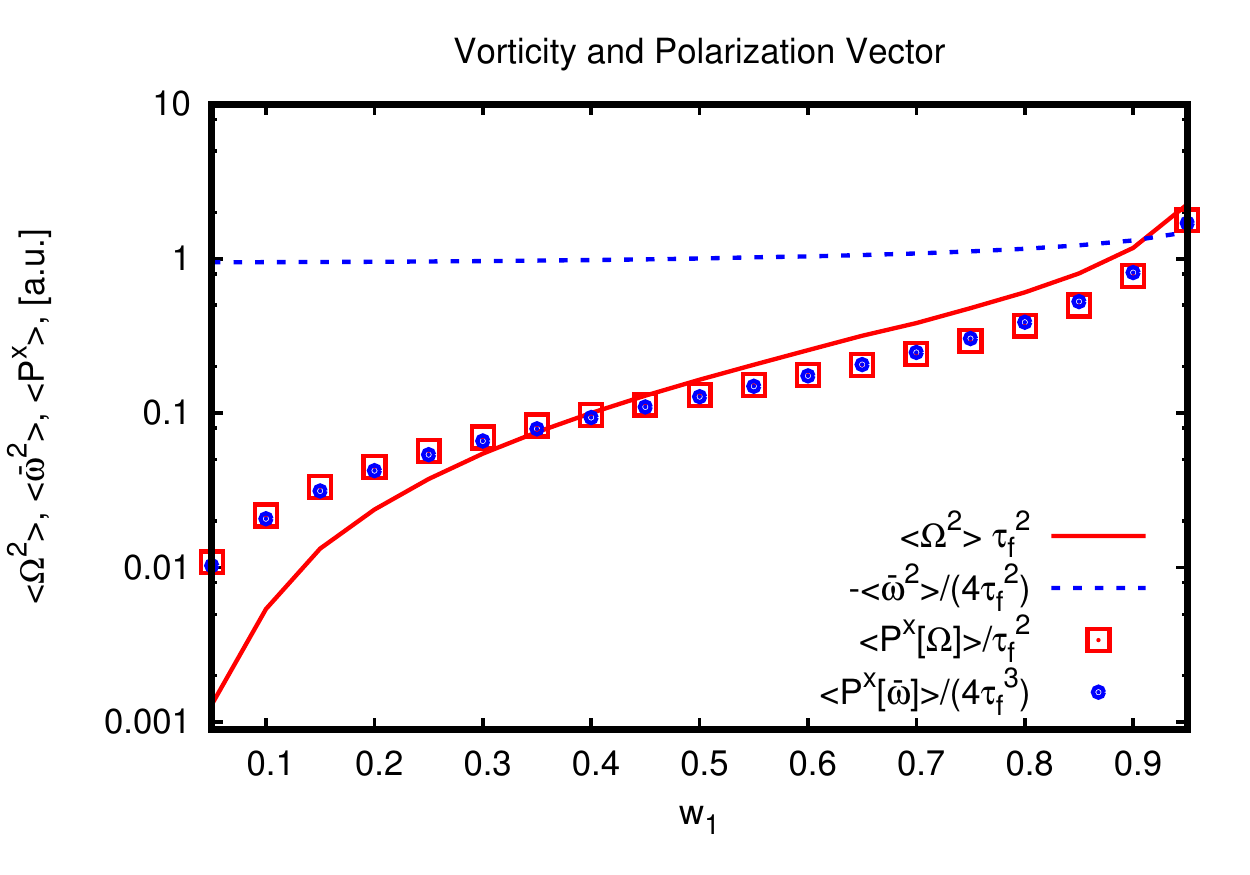}
  \includegraphics[width=0.47\linewidth]{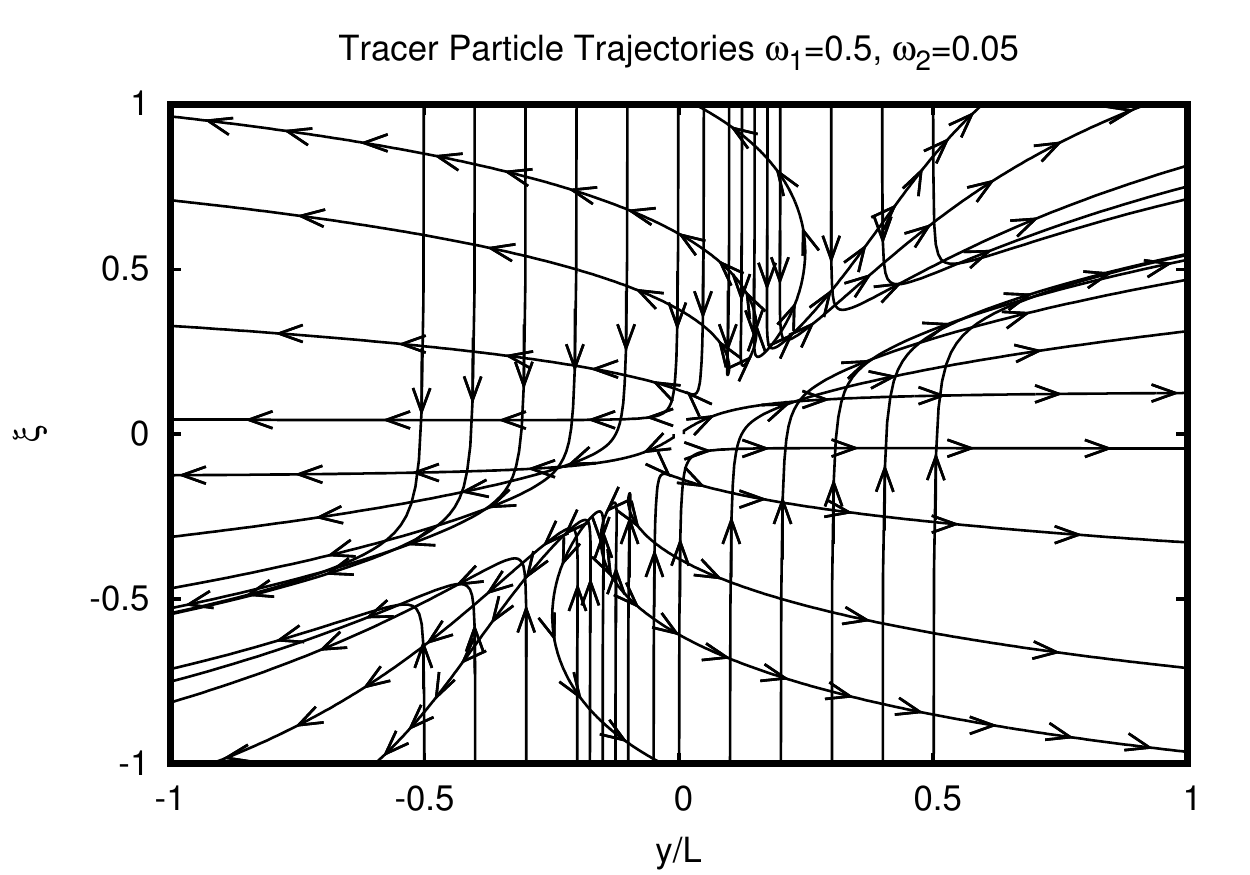}
  \caption{\label{fig:vort} Left: Average vorticity and polarization vector for $\omega_2=0.05$ as a function of angular frequency $\omega_1$ for isochronous decoupling at $\tau_f=10 L$ (results are essentially insensitive to the choice of $\tau_f$). Right: streamlines of tracer particles in the $y-\xi$ plane for $\omega_1=0.5,\omega_2=0.05$ for various initial locations $x,y,\xi$ at $\tau=0.01 L$. }
\end{figure*}

Results for flow coefficients $v_n$ are shown in Fig.~\ref{fig:coeff}, where values for $\omega_{1,2}$ in Fig.~\ref{fig:coeff} were chosen for qualitative illustration of Eq.~(\ref{eq:oursol}) rather than aiming for reproducing experimental data.  We find sizable direct, elliptic, and quadrupolar flow coefficients $v_1,v_2,v_4$, which exhibit both momentum and rapidity dependence in qualitatively agreement with relativistic heavy-ion experiment \cite{Back:2004je,Abelev:2013cva,Aad:2014vba},\footnote{Note that $v_1$ here corresponds to the rapidity-even flavor discussed in Ref.~\cite{Luzum:2010fb}. Furthermore, note that $p_\perp$ integrated results shown in Fig.~\ref{fig:coeff} are calculated with a $p_\perp$ cut of $p_\perp L>1$.}. Quantitatively, one finds that the analytic solution (\ref{eq:oursol}) leads to a rapidity distribution that is narrower than in heavy-ion experiments, but  consistent with numerical holography results for shock-wave collisions in Ref.~\cite{Casalderrey-Solana:2013aba}. 
Increasing $\omega_1$ does lead to somewhat wider rapidity profiles than those shown in Fig.~\ref{fig:coeff}, but only at the expense of an unrealistically large integrated $v_2$ coefficient, while increasing $\omega_2$ increases both $v_1,v_3$. 

\textbf{Angular Momentum and Vorticity}
Vorticity is a key characteristic of fluids and plays a fundamental role in first-principles derivations of relativistic fluid dynamics \cite{Baier:2007ix,Bhattacharyya:2008jc}.
Recently, a possible experimental handle on the fluid vorticity $X_{\mu\nu}$ in heavy-ion collisions using the average polarization of $\Lambda$ hyperons
\begin{equation}
\label{eq:pol}
\langle P^\mu\left[X\right]\rangle = \frac{1}{4 m_\Lambda} \frac{C^{\mu\nu \rho \sigma}\int d\Sigma^\lambda T_{\lambda \nu} X_{\rho \sigma}/T}{N_\Lambda} \,,
\end{equation}
has been suggested \cite{Liang:2004ph,Becattini:2007sr,Becattini:2013fla,Becattini:2015ska,Becattini:2016gvu,Pang:2016igs,STAR:2017ckg}. Here $T_{\lambda \nu}=\frac{\epsilon}{3}\left(4 u_\lambda u_\nu+g_{\lambda\nu}\right)$ is the energy-momentum tensor, $N_\Lambda$ is the momentum-integral of the $\Lambda$ hyperon spectrum (\ref{eq:fo}), and $C^{\mu\nu\rho\sigma}$ is the four-dimensional Levi-Civita symbol. There is some uncertainty as to which kind of vorticity should be used in the polarization $P^\mu$ above, candidates being the standard (kinematic) vorticity 
$X_{\mu\nu}=\Omega_{\mu\nu}$ or the ``thermal vorticity'' $X_{\mu\nu}=T \varpi_{\mu\nu}$\cite{Becattini:2016gvu} where
\begin{eqnarray}
  \Omega_{\mu\nu}&\equiv& \nabla^\perp_{[\mu}u_{\nu]}=
  \frac{1}{2 T}\left[\nabla_\mu (u_\nu T)-\nabla_\nu (u_\mu T)\right]\,,\nonumber\\
  \varpi_{\mu\nu}&=&\frac{1}{2}\left[\nabla_\mu \left(\frac{u_\nu}{T}\right)-\nabla_\nu\left( \frac{u_\mu}{T}\right)\right]\,.
  \end{eqnarray}
The average squared kinematic and thermal vorticity $\langle \left\{\Omega^2,\varpi^2\right\}\rangle\equiv \frac{\int d\Sigma^\tau \epsilon\, \left\{\Omega_{\mu\nu}\Omega^{\mu\nu},\varpi_{\mu\nu}\varpi^{\mu\nu}\right\}}{\int d\Sigma^\tau \epsilon}$ and associated polarization vector $\langle P^x[\left\{\Omega,\varpi\right\}]\rangle$ are shown in Fig.~\ref{fig:vort} as a function of angular rotation frequency $\omega_1$ for $\omega_2=0.05$.

As expected, average kinematic vorticity and the associated polarization vector exhibit qualitatively similar behavior  and become maximal for $\omega_1\rightarrow \pm 1$. Curiously, while $\langle P^\mu[\varpi]\rangle\simeq \langle P^\mu[\bar\Omega]\rangle$ up to an overall normalization, the average thermal vorticity does not vanish in the non-rotating limit $\omega_1\rightarrow 0$ for our analytic solution (\ref{eq:oursol}). An explanation for this behavior may be that while $\varpi$ combines effects of acceleration and kinematic vorticity, the effects of acceleration average to zero in $\langle P^\mu[\varpi]\rangle$ but not in $\langle \varpi^2\rangle$.

While non-vanishing vorticity and polarization vector can be expected for heavy-ion collisions on general grounds, one may ask if the system produced in heavy-ion collisions actually rotates in the sense of completing at least a single turn around its rotation axis. This question may be answered using Eq.~(\ref{eq:oursol}) by calculating the trajectories of tracer particles streaming with the velocity field (streamlines), defined as $\frac{d x^\mu}{d\tau}=\frac{1}{\gamma} u^\mu$. The streamline equations may be solved numerically using some initial condition $x^\mu$ and representative results are shown in Fig.~\ref{fig:vort}. While the streamlines do suggest sizable angular momentum of the system, not a single trajectory completes a full rotation. Thus, the system described by (\ref{eq:oursol}) does not rotate. Note, however, that for sufficiently high values of $\omega_2$ one does find local ``eddies'' in the transverse plane.

\textbf{Summary and Conclusions}
In this work, the phenomenological implications from studying analytic solutions for holographic heavy-ion collisions were studied. Considering off-center collisions of uncharged black holes in global AdS would lead to a deformed global Myers-Perry black hole  ringing down quickly via quasi-normal mode decay.  Employing previously derived solutions for the boundary stress tensor corresponding to a stationary Myers-Perry black hole in global AdS thus led to an analytic two-parameter family of ideal fluid dynamics solutions for the energy density and fluid velocity in Minkowski space given in Eq.~(\ref{eq:oursol}). These solutions constitute a qualitative model for the space-time evolution of an off-center heavy-ion collision, with full four-dimensional information available analytically.  (Note that for the parameter values and time scales relevant to the heavy ion phenomenology we focus on here, the superradiant instability of Myers-Perry-AdS does not play a significant role.) The analytic solutions (\ref{eq:oursol}) were found to exhibit large momentum anisotropies as well as vorticity, and may serve as a tool to investigate features of real-world heavy-ion collisions that are hard to extract from either numerical fluid dynamics solutions or experimental data. For instance, one of the questions that could be answered within the class of solutions (\ref{eq:oursol}) in this work was that the system, though possessing large angular momentum and vorticity, does not complete a single full rotation. Other uses for Eq.~(\ref{eq:oursol}) could be the study of initial state fluctuations on top of a smooth background along the lines of Refs.~\cite{Staig:2010pn,Gorda:2014msa,Hatta:2014jva}, or the effect of thermal fluctuations along the lines of Refs.~\cite{Kovtun:2011np,Murase:2013tma,Young:2013fka,Akamatsu:2016llw} or the onset of fluid turbulence along the lines of Refs.~\cite{1996NCimA.109.1405B,Romatschke:2007eb,Florchinger:2011qf}.  Furthermore, it would be instructive to revisit the analysis in the phenomenology section with data from full numerical simulations of off-center black hole collisions along the lines of Ref.~\cite{Bantilan:2014sra}.

We close by pointing out that the connection between Eq.~(\ref{eq:oursol}) and the gravitational description in terms of black holes in AdS space may lend itself as a concrete example to investigate if and how quantum informational tools, such as the spread of entanglement entropy or out-of-time-ordered correlation functions, get manifested in the real-time evolution of energy density and flow velocity in a heavy-ion collision, as well as suggest future potential experimental handles.

\textbf{Acknowledgments}
H. B. is supported by the European Research Council Grant No. ERC-2014-StG 639022-NewNGR, T.I. is supported by the Netherlands Organisation for Scientific Research (NWO) under the VIDI grant 680-47-518 and the Delta-Institute for Theoretical Physics ($\Delta$-ITP), which is funded by the Dutch Ministry of Education, Culture and Science (OCW) and P.R. is supported in part by the Department of Energy, DOE award No DE-SC0017905.   We would like to thank F.~Becattini, T.~Csörgő, P.~Figueras, Y.~Hatta, S.~Lin, J.~Nagle and J.~Noronha for fruitful discussions.

\onecolumngrid






\newpage
\section*{Supplemental Material}

In this supplemental material, details on obtaining the analytic solution (1) in the main text are presented. Let us consider the case of a large, uncharged, rotating black hole in global AdS space-time. Using Boyer-Lindquist coordinates $t^\prime,\tilde r,\tilde \theta,\phi^\prime,\chi^\prime$ the line element for five-dimensional global Kerr-AdS with rotations in $\phi^\prime,\chi^\prime$ parametrized by $|\omega_{1,2}|<1$, respectively, is given by Ref.~\cite{Hawking:1998kw,Hawking:1999dp,Gibbons:2004uw}:
\begin{align}
\frac{ds_5^2}{L^2} =& -\frac{\Delta(1+r^2)}{\Sigma_1 \Sigma_2} dt'^2 + \frac{\rho^2}{V-2M} d\tilde{r}^2 + \frac{2M}{\rho^2} \left( \frac{\Delta}{\Sigma_1 \Sigma_2} dt' - \frac{\omega_1 \sin^2 \tilde{\theta}}{\Sigma_1} d\phi' - \frac{\omega_2 \cos^2 \tilde{\theta}}{\Sigma_2} d\chi' \right)^2 \nonumber \\
&+\frac{\rho^2}{\Delta} d\tilde{\theta}^2 + \frac{r^2 + \omega_1^2}{\Sigma_1} \sin^2 \tilde{\theta} d\phi'^2 + \frac{r^2 + \omega_2^2}{\Sigma_2} \cos^2 \tilde{\theta} d\chi'^2\,,
\label{kerrads5_metric_bl}
\end{align}
where $L$ is the AdS radius, $M$ is the mass of the black hole, and shorthand notations
\begin{align}
V &\equiv \frac{(1+\tilde{r}^2)(\tilde{r}^2+\omega_1^2)(\tilde{r}^2+\omega_2^2)}{\tilde{r}^2}\,, \nonumber \\
\rho^2 &\equiv r^2 + \omega_1^2 \cos^2 \tilde{\theta} + \omega_2^2 \sin^2 \tilde{\theta}\,, \nonumber \\
\Delta &\equiv 1 - \omega_1^2 \cos^2 \tilde{\theta} - \omega_2^2 \sin^2 \tilde{\theta}\,, \nonumber \\
\Sigma_1 &\equiv 1-\omega_1^2, \quad \Sigma_2 \equiv 1-\omega_2^2\,,
\end{align}
where used. The above form is written in spheroidal coordinates. To have the boundary as static $S^3\times \mathbb{R}^1$ space-time, we introduce new coordinates $(r',\theta')$ by
\begin{align}
(1-\omega_1^2)r'^2 \sin^2 \theta' &= (\tilde{r}^2+\omega_1^2) \sin^2\tilde{\theta}, \\
(1-\omega_2^2)r'^2 \cos^2 \theta' &= (\tilde{r}^2+\omega_2^2) \cos^2\tilde{\theta}.
\end{align}
The full expression of the line element in the new coordinates is too cumbersome and we do not reproduce it here.
The boundary of the asymptotically AdS space-time is located at $r' \to \infty$, and near this boundary, the line element reads
\begin{align}
\frac{ds_5^2}{L^2} \simeq & -(1+r'^2) dt'^2 + \frac{dr'^2}{1+r'^2} + r'^2 \left(d\theta'^2 + \sin^2 \theta' \, d\phi'^2 + \cos^2 \theta' d\chi'^2 \right) \nonumber \\
&+ \frac{2M}{\Delta'^3 r'^2} dt'^2 + \frac{2M}{\Delta'^2 r'^6} dr'^2 +  \frac{2M \omega_1^2 \sin \theta'^4}{\Delta'^3 r'^2} d\phi'^2 + \frac{2M \omega_2^2 \cos \theta'^4}{\Delta'^3 r'^2} d\chi'^2 \nonumber \\
&- \frac{4M \omega_1 \sin \theta'^2}{\Delta'^3 r'^2} dt'd\phi'  - \frac{4M \omega_2 \cos \theta'^2}{\Delta'^3 r'^2} dt'd\chi' + \frac{4M \omega_1 \omega_2 \sin \theta'^2 \cos \theta'^2}{\Delta'^3 r'^2} d\phi' d\chi'\,,
\end{align}
where $\Delta' \equiv 1 - \omega_1^2 \cos^2 \theta' - \omega_2^2 \sin^2 \theta' \, (\equiv\gamma^{-2})$.
The leading behavior of the above line element on the $S^3\times\mathbb{R}^1$ boundary space-time with coordinates $x^a=\left(t^\prime,\theta^\prime,\phi^\prime,\chi^\prime\right)$ is given by 
\begin{equation}
ds^2_4 = L^2 \left(-dt'^2 + d\theta'^2 + \sin^2 \theta' \, d\phi'^2 + \cos^2 \theta' d\chi'^2 \right),
\end{equation}
and the conformal boundary energy-momentum tensor $T^{ab}$ may be obtained from the subleading terms as \cite{Bhattacharyya:2007vs}
\begin{equation}
  T^{ab}=\frac{\gamma^6 T_0^4}{3 L^2}\left(\begin{tabular}{cccc}
    $3+v^2$ & $0$ & $4 \omega_1$ & $4 \omega_2$\\
    $0$     & $1-v^2$ & $0$ & $0$\\
    $4 \omega_1$ & $0$ & $3\omega_1^2+{\rm csc}^2\theta^\prime-\omega_2^2 \cot^2\theta^\prime$ & $4 \omega_1 \omega_2$\\
    $4\omega_2$ & $0$ & $4 \omega_1 \omega_2$ & $3 \omega_2^2+{\rm sec}^2 \theta^\prime - \omega_1^2 \tan^2\theta^\prime$
  \end{tabular}\right)\,,
\end{equation}
where $v=\sqrt{\omega_1^2 \sin^2 \theta^\prime+\omega_2^2 \cos^2\theta^\prime}$, $\gamma=\frac{1}{\sqrt{1-v^2}}$ and $T_0=\left(\frac{3 M}{8 \pi G_5 L}\right)^{1/4}$ sets the overall energy scale with $G_5$ being five-dimensional Newton's constant. This energy-momentum tensor corresponds to the case of an uncharged conformal fluid on $S^3\times \mathbb{R}^1$ which is rigidly rotating about two axes with angular velocities $\omega_1,\omega_2$ with $|\omega_{1,2}|<1$, respectively. Since the time-evolution of this system is stationary, the energy-momentum tensor corresponds to a solution of the equations of motion of ideal fluid dynamics with energy density $\epsilon_{S^3\times\mathbb{R}^1}$ and fluid velocity $u^a_{S^3\times\mathbb{R}^1}$ defined as the time-like eigenvector and eigenvalue of the energy-momentum tensor through $u_b T^{ab}=-\epsilon  u^a$:
\begin{equation}
  \epsilon_{S^3\times \mathbb{R}^1}=\frac{4 T_0^4}{\left(2 - \omega_1^2-\omega_2^2+(\omega_1^2-\omega_2^2)\cos 2 \theta^\prime\right)^2}\,,\quad
  u^a_{S^3\times \mathbb{R}^1}=\frac{\gamma}{L} \left(1,0,\omega_1,\omega_2\right)\,.
  \end{equation}
The coordinate transformation from $S^3\times \mathbb{R}^1$ to Minkowski space-time in polar coordinates $t,r,\theta,\phi$
is straightforward and given by 
\begin{eqnarray}
  \label{eq:coo-tra}
  t^\prime&=&{\rm arctan}\frac{2 L t}{L^2+r^2-t^2}\,,\nonumber\\
  \theta^\prime &=&{\rm arctan} \frac{2 L r \sin\theta}{\sqrt{L^4+2 L^2 t^2+(r^2-t^2)^2+2 L^2 r^2 \cos (2 \theta)}}\,,\\
  \phi^\prime &=& \phi\,,\nonumber\\
  \chi^\prime &=&-{\rm arctan} \frac{L^2 -r^2+t^2}{2 L r \cos\theta}\,.\nonumber
\end{eqnarray}
The above coordinate transformation leads to a line element for the boundary space-time
\begin{equation}
  ds^2=W^{-2}\left(-dt^2+dr^2+r^2 d\theta^2+r^2 \sin^2\theta d^2\phi\right)\,,\quad
  W^2=\frac{L^4+(r^2-t^2)^2+2 L^2 (r^2+t^2)}{4 L^4}\,,
  \end{equation}
which up to a conformal factor of $W^2$ is that of Minkowski space-time in polar coordinates. The coordinate-transformation together with a Weyl rescaling of the metric to remove the conformal factor in $ds^2$ leads to an energy-density and fluid velocity in Minkowski space-time $\mathbb{R}^{3,1}$ given by 
\begin{eqnarray}
\label{eq:minksol}
  \epsilon_{\mathbb{R}^{3,1}}&=&\frac{16 L^8 T_0^4}{\left[(L^4+(r^2-t^2)^2+2 L^2 t^2)(1-\omega_2^2)+2 L^2 r^2 (1-\omega_1^2)+2 L^2 r^2 (\omega_1^2-\omega_2^2)\cos2\theta\right]^2}\,,\nonumber\\
  u^t_{\mathbb{R}^{3,1}}&=& \frac{L^2+r^2+2 r t \omega_2 \cos\theta}{\sqrt{(L^4+(r^2-t^2)^2+2 L^2 t^2)(1-\omega_2^2)+2 L^2 r^2 (1-\omega_1^2)+2 L^2 r^2 (\omega_1^2-\omega_2^2)\cos2\theta}}\,,\nonumber\\
  \frac{u^r}{u^t}_{\mathbb{R}^{3,1}}&=&\frac{2 r t+(L^2+r^2+t^2) \omega_2 \cos \theta}{L^2+r^2+2 r t \omega_2 \cos\theta}\,,\nonumber\\
 \frac{u^\theta}{u^t}_{\mathbb{R}^{3,1}}&=&\frac{-(L^2-r^2+t^2)\omega_2 \sin\theta}{r (L^2+r^2+2 r t \omega_2 \cos\theta)}\,,\nonumber\\
 \frac{u^\phi}{u^t}_{\mathbb{R}^{3,1}}&=&\frac{2 L \omega_1}{L^2+r^2+2 r t \omega_2 \cos\theta}\,.
  \end{eqnarray}
Note that the conformal transformation leads to an extra factor of $W^{-4}$ in the scaling of the energy density from $S^{3}\times \mathbb{R}^1$ to Minkowski space-time, cf. the discussion in Ref.~\cite{Bantilan:2012vu}. 

In a final step, we transform from Minkowski polar coordinates to Milne coordinates $x^\alpha=\left(\tau,x_1,x_2,\xi\right)$ with $\tau=\sqrt{t^2-r^2\sin^2\theta\sin^2\phi}$, $x_1=r \cos \theta$, $x_2=r \sin\theta \cos\phi$, $\xi={\rm arctanh}\frac{r \sin \theta\sin\phi}{t}$. We find $g_{\alpha\beta}={\rm diag}\left(-1,1,1,\tau^2\right)$ and
\begin{eqnarray}
\label{eq:genres}
  \epsilon &=&\frac{16 L^8 T_0^4}{\left[(L^4+2 L^2 {\bf x}_\perp^2+(\tau^2-{\bf x}_\perp^2)^2)(1-\omega_2^2)+2 L^2 (\tau^2-2 x_2^2)(\omega_1^2-\omega_2^2)+2 L^2 \tau^2(1-\omega_1^2)\cosh 2\xi\right]^2}\,,\nonumber\\
  u^\tau&=&\left[(L^2+\tau^2+{\bf x}_\perp^2)\cosh \xi+2 (\tau \omega_2 x_1-L \omega_1 x_2 \sinh\xi)\right]\left(\frac{\epsilon}{16 L^8 T_0^4}\right)^{1/4}
  \nonumber\\
  \frac{u^1}{u^\tau}&=&\frac{2 \tau x_1 \cosh\xi+\omega_2(L^2+\tau^2+x_1^2-x_2^2)}{(L^2+\tau^2+{\bf x}_\perp^2)\cosh \xi+2 (\tau \omega_2 x_1-L \omega_1 x_2 \sinh\xi)}\,,\nonumber\\
   \frac{u^2}{u^\tau}&=&\frac{2 \tau x_2 \cosh\xi+2 \omega_2 x_1 x_2-2 L \tau \omega_1 \sinh\xi}{(L^2+\tau^2+{\bf x}_\perp^2)\cosh \xi+2 (\tau \omega_2 x_1-L \omega_1 x_2 \sinh\xi)}\,,\nonumber\\
 \frac{u^\xi}{u^\tau}&=&-\frac{(L^2-\tau^2+{\bf x}_\perp^2)\sinh\xi-2 L \omega_1 x_2 \cosh\xi}{\tau\left[(L^2+\tau^2+{\bf x}_\perp^2)\cosh \xi+2 (\tau \omega_2 x_1-L \omega_1 x_2 \sinh\xi)\right]}\,,
  \end{eqnarray}
where ${\bf x}_\perp^2=x_1^2+x_2^2$, which is Eq.~(1) in the main text. It is straightforward to check that equations (\ref{eq:genres}) fulfill the ideal hydrodynamic equations of motion.



\bibliographystyle{hunsrt}
\bibliography{pp-hydro}

\end{document}